\documentclass[11pt]{article}
\usepackage[utf8]{inputenc}
\usepackage[english]{babel} 
\usepackage{sao1}
\usepackage{epsfig}

\begin{document}
\title{On the Discovery of  Strong Magnetic Field in HD\,34736 Binary System\footnote{The work is based on the observations obtained with the 6-m telescope of the SAO RAS.}}
\author{E.~A.~ Semenko\footnote{E-mail: sea@sao.ru}, I.~I.~Romanyuk, D.~O.~Kudryavtsev, I.~A.~Yakunin}
\institute{\saoname}
\maketitle

\begin{abstract}
We present  the results of a study of   the star HD\,34736. The
spectropolarimetric observations  carried out at the 6-m telescope
showed the presence of a strong variable longitudinal magnetic
field, exceeding $-4500$~G. The analysis of the HIPPARCOS
photometry gives a set of possible periods of the brightness
variability of the star, of which $0\fd3603$ is   preferred. The
variable radial velocity of spectral lines  of the star and some
signatures of lines of at least one other component show that
HD\,34736 is a double short-period system. Modeling of the spectra
allowed us to estimate the effective temperature
$T_\mathrm{eff}$~of the stars ($13\,700$ and $11\,500$~K) and
their projected rotational velocities $v\sin i$ ($73$ and
$\geq90$~km\,s$^{-1}$). The analysis of all the available
information about the star allows us to hypothesize that the
object of study is a close, possibly interacting binary system.
\keywords{stars: magnetic field---binaries: close---stars:
fundamental parameters---stars: chemically peculiar---stars:
individual: HD\,34736}
\end{abstract}

\section{INTRODUCTION}
The study of magnetic properties of early-type stars provides a
unique opportunity to study in detail the features of the
processes of their evolution, formation and variation of the
magnetic field strength with age.  This is primarily due  to the
rapid evolution of massive stars. A major part of B-type stars is
found in open clusters and associations, and hence the accuracy of
determination of their age is significantly higher than for the
field stars. All these factors have facilitated the start  of a
 large observational program called {\it Magnetic Fields of Massive Stars} at the  Special Astrophysical Observatory of Russian Academy of Sciences (SAO RAS).  The conceptual issues
of the  program, formulating the problem,  and selecting the
objects of study are presented in the paper by Romanyuk  and
Yakunin~\cite{romanyuk_yakunin:2012}. Spectropolarimetric observations with the 6-m BTA telescope within this program started at the end of 2010~and are still ongoing.
Recently, our research has been focused on the study of the
already known and the search for new stars with magnetic fields in
the nearby  Ori\,OB1 association. In the course of the survey of
chemically peculiar (CP) stars of the association, the star
HD\,35298 was studied in detail~\cite{yakunin:2013}. The observations of
poorly-studied CP stars of the association allowed us to detect
the signs of   magnetic field in four stars. This paper is devoted
to the features of one of them, HD\,34736.

A chemically peculiar star HD\,34736 ($m_V=7\fm82$)  with silicon abundance
anomalies~\cite{roman:1978} is a member of the
Ori\,OB1 association (sub-group~C)~\cite{brown_etal:1994}. A large amount of the
photometric   data (the systems
$UBV$~\cite{deutschman_etal:1976},
Str\"omgren~\cite{vogt_faundez:1979}, Maitzen~$\Delta
a$~\cite{maitzen_vogt:1983},
Walraven~\cite{de-geus_etal:1990}) provides confident
estimates of the main characteristics of the star. According to
the Geneva photometry indices, North and
Kramer~\cite{north_cramer:1984} predict the surface
magnetic field of the star of ~\mbox{$B_\mathrm{s}\approx1.9$~kG.}
However,  the literature contains   no evidence on the direct
measurement of this magnetic field.

Based on the analysis of data  presented in the work of Brown et
al.~\cite{brown_etal:1994}, the catalog of CP  stars by
Renson and Manfroid~\cite{renson_manfroid:2009}, and
other literary sources, we have selected the star HD\,34736   as a
candidate for the spectropolarimetric observations with the 6-m
telescope of the SAO RAS~\cite{romanyuk_etal:2013}.
Three observations of the star were performed to date, their
description is contained in Section~2. The results of measurements
of the Zeeman effect in the spectra of the star and determining
the magnitude of its magnetic field are shown in Section~3.
Section 4 is devoted to the estimation of the main physical
parameters of HD\,34736. The final Section 5 discusses the results
we have obtained.

\section{SPECTROSCOPIC  OBSERVATIONS \\  AND PROCESSING}
For the first time the star HD\,34736 was observed by us on the
6-m telescope in October 2013.  In December of the same year, two
more sets of spectra were obtained. An excerpt from the log of
observations is presented in Table~\ref{table:1}. It
provides the data on the Julian date (HJD) at the time of
mid-exposure, the spectral range, and the signal-to-noise ratio of
the registered spectra of polarized light.  The observations were
carried out on the Main Stellar Spectrograph~(MSS) of the BTA, the
description of the optical scheme of which and its instrumental
capabilities can be found on the web page of the
device.\footnote{\tt http://www.sao.ru/hq/lizm/mss/en/} The
spectrograph was used in the spectropolarimetric mode using a
circular polarization analyzer and a slit unit with a double image
slicer~\cite{chountonov:2004}. The polarization analyzer
has a $\lambda/4$ phase plate, which has two fixed positions at
90\degr\ relative to each other. The characteristic property of
this analyzer is that the resulting spectrum of circularly
polarized
 light of a star should be appropriately
 calculated as the average of  two exposures at different
positions of the phase plate. The averaging allows  to effectively
get rid of residual traces of cosmic particles and also to remove
the possible instrumental polarization, as the spectra of
oppositely oriented polarized light are registered at the same
regions of the detector. In all the cases, the spectra were
registered using the E2V~CCD\,42-90 chip sized $2048\times4600$
elements. The average inverse linear dispersion of the recorded
spectra is about 0.1215~\AA\ per pixel.

\begin{table}[]
\caption{Log of observations of HD\,34736 and standard stars}
\label{table:1}
\begin{center}
\medskip
\begin{tabular}{c| c| c| r}
\hline
\multicolumn{1}{c|}{Star}    &  HJD\,2450000+ & $\lambda$, {\AA}           &  $S/N$ \\
\hline
HD\,34736  & 6589.4929       & 4402.6--4955.8          & 250 \\
53\,Cam    & 6589.5569       &                         & 270 \\
$o$\,UMa   & 6589.5578       &                         & 690 \\
HD\,34736  & 6639.4980       & 4428.3--4983.7          & 220 \\
$\alpha^2$\,CVn & 6639.6643     &                         & 1150 \\
HD\,33256  &  6639.4717      &                         & 270 \\
HD\,34736  & 6644.4389       & 4427.3--4982.3          & 300 \\
53\,Cam    & 6644.5097       &                         & 300 \\
$o$\,UMa   & 6644.5171       &                         & 830 \\
\hline
\end{tabular}
\end{center}
\end{table}

A set of calibration images needed to perform the extraction of
spectral data was typical for such problems and included the
images of zero exposure frames~(bias), spectra of the continuous
spectrum  source for the subsequent flat-fielding, and a
comparison spectrum, for which we have used the emission spectrum
of the Th-Ar lamp.

All the procedures related to the processing and extraction of
one-dimensional spectra were ran in the ESO-MIDAS using the LONG
context and a set of programs written by
D.~Kudryavtsev~\cite{kudryavtsev:2000}. The spectra were
normalized to the continuum level, and the wavelength scale was
corrected for the orbital motion of  Earth.

\section{MAGNETIC FIELD OF HD\,34736}
\begin{figure*}[!h]
\vspace{2mm}
\centering\includegraphics[width=10cm]{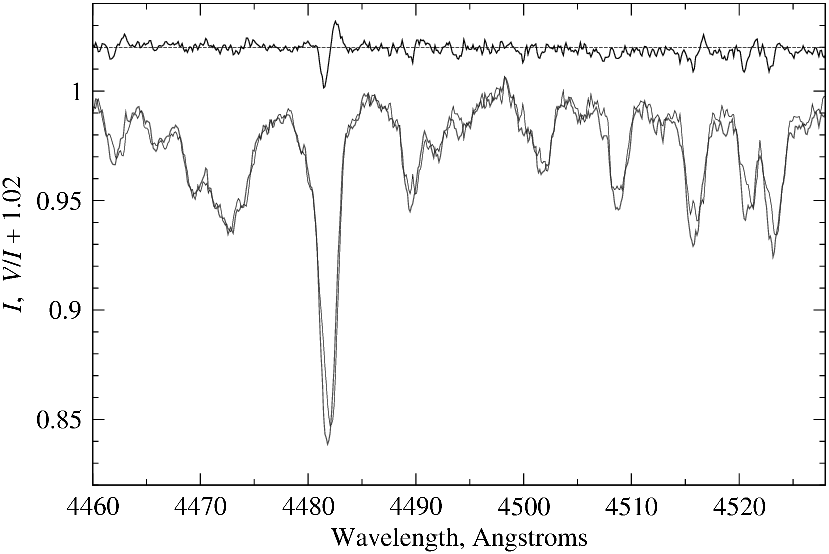}
\caption{The right-hand and   left-hand   circularly polarized
spectra of the star HD\,34736 in the region of the
Mg\,II~4481~\AA\ line. The normalized Stokes~$V$ parameter,
shifted along the $y$-axis, is given on the top.}
\label{figure:1}
\end{figure*}

\begin{table*}
\caption{The results of determination of the longitudinal
magnetic field of  HD\,34736 and standard stars. Columns 3--5
represent  the results obtained by measuring the centers of
gravity of spectral lines,  approximating function minima and by
the linear regression method ~\cite{bagnulo_etal:2002},
respectively. The number of measured lines is indicated in the
parentheses} \label{table:2}
\medskip
\begin{center}
\begin{tabular}{c| c| r@{$\,\pm\,$}r c | r@{$\,\pm\,$}r c | r@{$\,\pm\,$}l | l}
\hline
\multicolumn{1}{c|}{Star} & HJD\,2450000+ & \multicolumn{3}{c|}{$B_\mathrm{e}(\mathrm{COG})\pm\sigma$, G} & \multicolumn{3}{c|}{$B_\mathrm{e}(\textrm{AF})\pm\sigma$, G} & \multicolumn{2}{c|}{$B_\mathrm{e}(\textrm{regres})\pm\sigma$, G}& Sp \\
\hline
HD\,34736  & 6589.4929       & $-3500$ & $440$& (36)   & $-4380$ & $720$ & (13)  & ~~~~$-2147$& $135$ & B9 \\
53\,Cam    & 6589.5569       & $740$ & $70$   &(228)     & $690$ & $80$  &(203)    & $935$ & $\phantom{1}50$ & A2p \\
$o$\,UMa   & 6589.5578       & $-59$ & $50$   &(335)     & $-80$ & $60$  &(328)    & $-80$ & $\phantom{1}50$ & G5 \\
HD\,34736  & 6639.4980       & $-160$ & $530$ & (33)    & $-160$ & $1170$& (12)  & $782$ & $170$  & B9 \\
$\alpha^2$\,CVn & 6639.6643  & $-680$ & $50$  &(242)    & $-760$ & $50$  &(159)  & $-710$ & $\phantom{1}50$  & A0 \\
HD\,33256  & 6639.4717       & $-33$ & $50$   &(205) & $-30$ & $50$      &(188) & $-50$ & $\phantom{1}50$    & F2 \\
HD\,34736  & 6644.4389       & $-4580$ & $560$& (45) & $-5080$ & $1040$  & (16) & $-3790$ & $140$ & B9 \\
53\,Cam    & 6644.5097       & $-3340$ & $110$&(193) & $-3460$ & $120$   &(169) & $-2930$ & $\phantom{1}50$  & A2p\\
$o$\,UMa   & 6644.5171       & $-10$ & $60$   &(266)    & $-10$ & $60$   &(262)    & $-10$ & $\phantom{1}50$ & G5 \\
\hline
\end{tabular}
\end{center}
\end{table*}

The spectra of HD\,34736  are characterized by complex profiles of
most of the lines (Fig.~\ref{figure:1}): the same lines
appear differently in the spectra of right and left-hand circular
polarizations. This phenomenon indicates the complex geometry of
the global magnetic field of the star and its considerable
strength. The stellar spectrum relatively poor in lines   is a
consequence of two factors: high effective
temperature~($13\,800$~K, according
to~\cite{brown_etal:1994}) and rapid rotation of the
star. Our experience shows that in case of spectra with complex
line profiles, different longitudinal field methods show different
results (see, e.g.~\cite{yakunin:2013}). Sometimes these
differences prove to be statistically significant. In order to
maximize the accuracy of our conclusions, we have at the same time
used the method for measuring the Zeeman effect consisting in
finding the shifts between the line profiles in the spectra of the
left and right-hand circular polarizations, and the method of
longitudinal field estimation proposed by Bagnulo et
al.~\cite{bagnulo_etal:2002}. In practice, we can take
several different quantities for the spectral line wavelength. For
example, the wavelength of the center of gravity of a profile or
the wavelength of the   approximating function minimum when it
comes to absorption lines.  The Gauss function  is most commonly
used for the latter. It is easy to understand that in the case of
complex line profiles, distorted, for example, by the crossover
effect, the wavelength of the spectral line center will be
different. This discrepancy affects the results and is especially
critical when we try to measure the weak $B_\mathrm{e}$ fields.

Notwithstanding the fact that the process of obtaining and
processing of the spectral data that we implement on the 6-m
telescope is aimed at the maximum account of the instrumental
effects, we perform additional verification of the results of our
measurements. Along with the studied stars,  we obtained the
spectra of standard stars every night. To verify the absence of
instrumental shifts of spectral lines, we use the stars of late
spectral classes with a big number of  narrow lines---the
so-called zero-field standards. The second group of standards
consists of those  CP stars, the magnitudes and polarities of
whose magnetic fields are well known. In the observations
described we used the standard stars 53\,Cam, $\alpha^2$\,CVn
(magnetic) and HD\,33256, $o$\,UMa (zero-field standards).

The summary Table~\ref{table:2} contains the results of
the longitudinal magnetic field measurements for the studied star
and standard stars. Columns 3~and~4 in parentheses indicate the
number of measured lines.

A comparative analysis of the   Table~\ref{table:2} data
allows us to confidently infer the presence of a magnetic field on
the surface of HD\,34736. At one night the longitudinal field of
the star was measured close to zero; however, there is a crossover
effect (the difference in the shapes of lines of the right-and
left-hand circularly polarized spectra) which happens when the
star is observed from the magnetic equator. All the three
measurements of the stars without magnetic fields gave a zero
result within the measurement errors, thus confirming the absence
of significant instrumental effects. The measured longitudinal
field of the 53\,Cam and $\alpha^2$\,CVn   stars is in good
agreement with the expected values, presented in the form of
dependencies, for instance, in~\cite{wade_etal:2000}.

As for the accuracy of  measurements, it is worth noting the
following. Two methods for determining the field based on the
measurement of  shifts of the  polarized components of   spectral
lines give a result comparable in accuracy  (columns 3~and~4 of
Table~\ref{table:2}). The linear regression method
(column 5), described in~\cite{bagnulo_etal:2002}, has
some restrictions. The most significant restriction concerns the
weak field mode---the method is applicable when the influence of
the Zeeman effect on the shape of spectral line profiles is
significantly weaker than the other broadening mechanisms. In the
case of the star HD\,34736, where the projected rotational
velocity amounts to   73~km\,s$^{-1}$, the weak field
approximation will be valid up to the longitudinal field values of
\mbox{4--5~kG}, depending on the lines. This means that the
$B_\mathrm{e}$ values  from column 5 are valid. The second
restriction of the linear regression method is due to the quality
of spectrum normalization to the continuum level. Since the
normalization cannot always be done quite accurately in all parts
of the spectrum, we believe that the errors can be somewhat
underestimated for some measurements of magnetic stars indicated
in column 5.

\section{PHYSICAL PARAMETERS OF THE STAR}
In the literature one can find some estimates of the effective
temperature and luminosity of the star. Our independent estimates of
$T_\mathrm{eff}$ agree well with the data of
Brown et al~\cite{brown_etal:1994}~(13\,800 K) and
Glagolevskij~\cite{glagolevskij:1994}~(12\,800 K).

We have determined  the effective temperature of the star and its
surface gravity with the use of  the existing calibrations for the
photometric indices in the Str\"omgren   and Geneva systems.

The index values of the first system are taken
from~\cite{deutschman_etal:1976}
and~\cite{vogt_faundez:1979}. The calibrations from
Napiwotzki et al.~\cite{napiwotzki_etal:1993},
implemented in the  {\tt uvbybeta\_new} code, give two effective
temperature values:  $13\,620$~K  and $13\,180$~K. The latter
corresponds to the calibration of the photometric   index $[u-b]$,
used basically  for the stars with   \mbox{$T_\mathrm{eff} >
9500$}~K. The $\log g$ value  is equal to $4.31$, according to
Napiwotzki et al.~\cite{napiwotzki_etal:1993}. The
application of dependences to the photometric indices described
in~\cite{moon_dworetsky:1985} gives
$T_\mathrm{eff}=13\,756$~K, $\log g=4.21$.

For the photometric indices of the Geneva   system, we have used
the calibrations from~\cite{kuenzli_etal:1997}. The
given method for determining the effective temperature is
sensitive to the interstellar extinction, which, accounting for
the far distance of the star
\mbox{($\pi=1.78\pm0.94$~mas~\cite{vanleeuwen:2007})}
and its position in the young association, can reach significant
values.  Since we have   spectra  in our disposal with only a
limited wavelength range, we attempted to estimate the
\mbox{$E(B-V)$} color excess from the   photometry. The
temperature calibrations by Moon and
Dworetsky~\cite{moon_dworetsky:1985} for the $uvby$
photometric system  allow us to estimate the color excess
\mbox{$E(b-y)=0\fm012$}. Then we find \mbox{$E(B-V)=0\fm017$} and
$E(B2-V1)=0\fm015$, using the $E(B-V)=1.43\,E(b-y)=$
$1.14\,E(B2-V1)$ relations that link the $UBV$, Str\"om\-gren, and
Geneva photometric systems. Another method to estimate the
interstellar absorption values is the $UBV$ photometry in case the
\mbox{$(U-B)$} and \mbox{$(B-V)$} indices are known. For HD\,34736
the relationship \mbox{$E(B-V)=(B-V)-0.332\,Q$} is true, where the
$Q$ parameter is defined as
$$Q=(U-B)-0.72\,(B-V)-0.05\,(B-V)^2.$$
This method gives a significantly higher value of
\mbox{$E(B-V)=0\fm047$}.  A similar result is obtained if we
follow the work of Brown et al.~\cite{brown_etal:1994},
where all the data were obtained from the photometric indices of
the Walraven system: \mbox{$E(B-V)=0\fm038$}. For the further
calculations we took the average of the last two quantities:
\mbox{$E(B-V)=0\fm043$}. The interstellar absorption in
the $V$-band is\linebreak \mbox{$A_V=3.11\,E(B-V)=0\fm134$}.

The GCPD   catalog of photometric data\footnote{\tt
http://obswww.unige.ch/gcpd/} contains information  on the
photometry of HD\,34736 in the Geneva system. After applying the
photometric calibrations from K\"unzli et
al.~\cite{kuenzli_etal:1997}, we have obtained two
parameter sets depending on the metallicity of the star:
$T_\mathrm{eff}=13\,500$~K, $\log g=4.13$ for $[{\rm M/H}]=0$, and
$T_\mathrm{eff}=13\,200$~K, $\log g=4.20$ for \mbox{$[{\rm
M/H}]=+1$}. The color excess  $E(B2-V1)=0\fm038$.

We can see that all the photometric estimates of the effective
temperature of the star are in good agreement, but a comparison of
the observed H$_{\beta}$  hydrogen line profile with the
calculated one testifies about a slightly higher temperature.
Therefore, we finally determine $T_\mathrm{eff}$ as
\mbox{$13\,700\pm400$~K}. The shape of the   H$_{\beta}$  line
  can be satisfactorily described in
the assumption that the surface gravity is equal to
\mbox{$4.15\pm0.15$~dex.} This value is only  $0.05$~dex smaller
than the average photometric estimate  of $\log g$.

\begin{figure}
\centering\includegraphics[width=10cm]{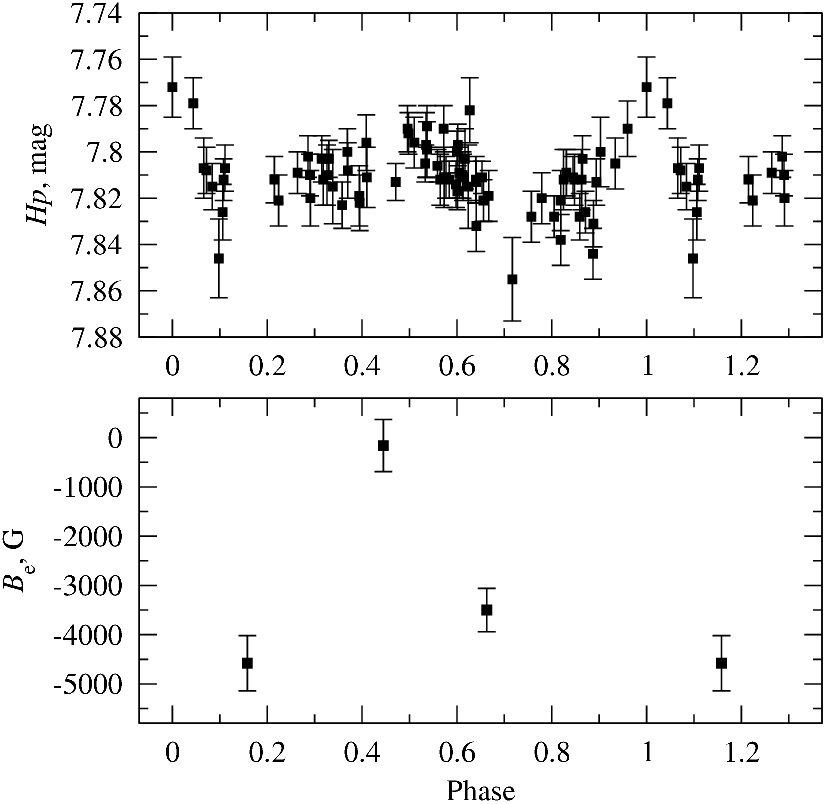}
\caption{Photometric variability of   HD\,34736 according to the
HIPPARCOS data with the period of $0\fd3603$ (top) and our
measurements of the longitudinal field $B_\mathrm{e}$ of the star,
phased with the same period (bottom).}
\label{figure:2}
\end{figure}

\begin{figure*}
\centering\includegraphics[width=15cm]{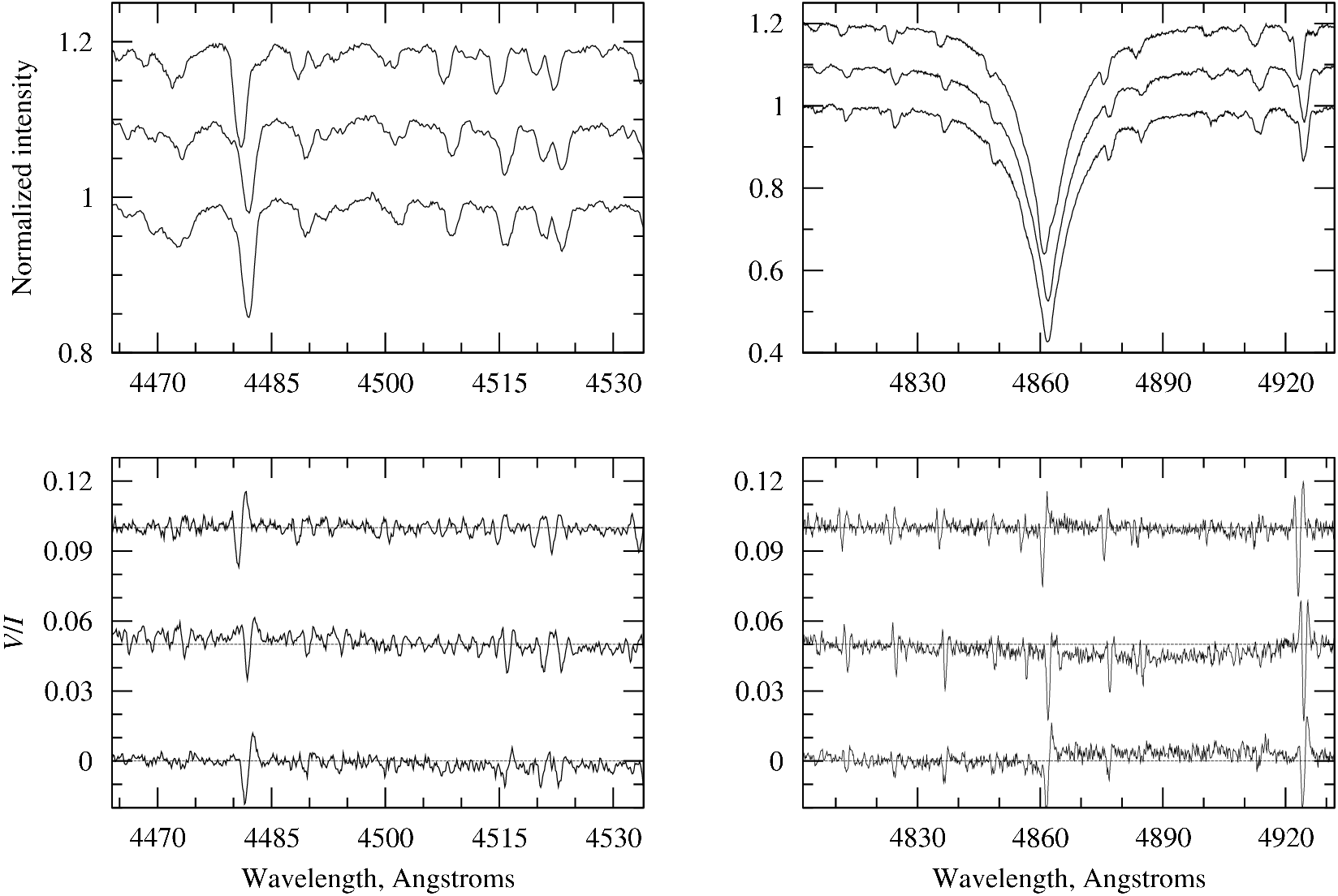}
\caption{Two regions in the spectrum of HD\,34736, demonstrating
the variability of the magnetic field and   radial velocity. The
upper graphs
 contain the intensity spectra $I$, normalized to the continuum, the bottom
graphs are the normalized spectra of the Stokes~$V$ parameters.
The spectra for different  rotation phases are given with the
shift along the  $y$ axis, from bottom to top: 0.15, 0.45, and
0.65. The phases are calculated in accordance with the period
$0\fd3603$.}
\label{figure:3}
\end{figure*}

The star HD\,34736 belongs to moderately fast rotators. From the
measurements of the profile of the Fe\,II~4508~\AA\ line, which is
the least sensitive to the magnetic field, we find \mbox{$v\sin
i=73\pm7$\,km\,s$^{-1}$}. The specified value implies the period
of stellar rotation shorter than three days. To determine the
period of stellar rotation, we attempted to analyze the photometry
data obtained by the HIPPARCOS mission and in the
ASAS survey~\cite{pojmanski:2002}. Using the Dimming
method~\cite{deeming:1975}, we can get a few values of
the likely period, from 0.3 to 1.6 days (e.g.,
Fig.~\ref{figure:2}). With some of them we can
satisfactorily reconcile the values of the longitudinal field of
the star. However, we have to bear in mind that   three
measurements of $B_\mathrm{e}$ are not enough to draw firm
conclusions about the period. In any case, only the further
spectropolarimetric observations of the star, given the
longitudinal field variability amplitude might yield the most
accurate idea about the variability of HD\,34736.

\begin{figure}
\centering\includegraphics[width=10cm]{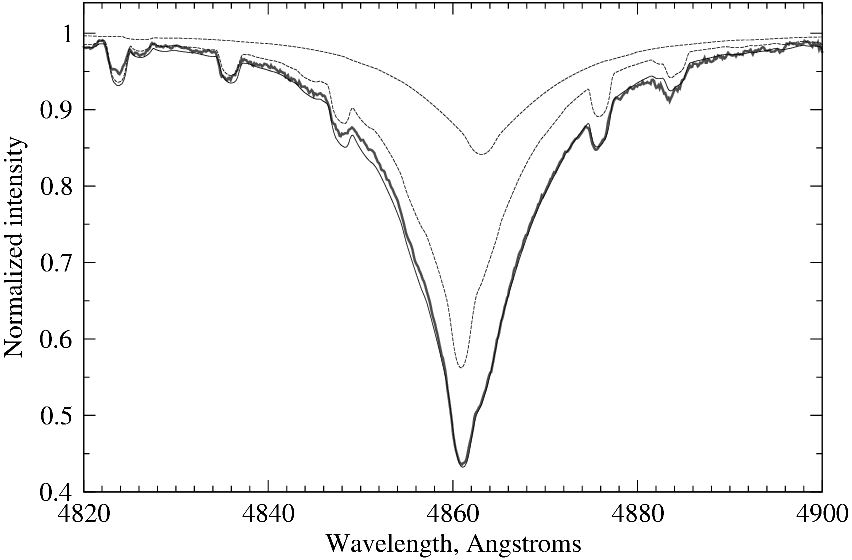}
\caption{The   spectrum of HD\,34736 in the region of the
H$_\beta$ hydrogen line with indications of the line of the second
component. The smooth thin black line is the synthetic spectrum
calculated with the parameters mentioned in the text. The dotted
lines correspond to individual spectra of the binary system.}
\label{figure:4}
\end{figure}

An important fact is a note in the HIPPARCOS catalog that the
variability of the star is caused by its duality. We get a direct
confirmation of the presence of another star from the analysis of
the spectra. Firstly, from the three spectra obtained by us,  the
radial velocity of the star significantly differs in two
cases~(Fig.~\ref{figure:3}):
\mbox{$+31\pm6$~km\,s$^{-1}$} and \mbox{$-41\pm5$~km\,s$^{-1}$}.
Secondly, the spectrum of \mbox{October 24, 2013} \linebreak
(HJD\,2456589.4929) in the region of  the H$_\beta$ line bears the
features of a hydrogen line belonging to the second star
(Fig.~\ref{figure:4}). The asymmetry manifests itself in
the majority of other lines too (Fe, Cr, Mg, and Si). We were able
to describe the H$_\beta$ line shape presenting it by a
combination of   two lines. The best agreement was reached in an
assumption  that the primary component of the system is
characterized by \mbox{$T_\mathrm{eff}=13\,700$~K} and \mbox{$\log
g=4.0$}, while the secondary component is somewhat cooler:
\mbox{$T_\mathrm{eff}=11\,500$~K} and \mbox{$\log g=4.0$}.  We
have earlier found the projected rotational velocity of the hotter
star  \mbox{($v\sin i = 73$~km\,s$^{-1}$)}. It is more difficult
to judge on the rotation of the second star. The wavelength range
of our spectra  has virtually no lines that could be confidently
attributed to the cooler star, except for the strongest line
Mg\,II 4481\,\AA. The degree of broadening of the
 Mg\,II and H$_\beta$ lines of the second component corresponds to the rotation of the
star at  $v\sin i$ value of not less than 90~km\,s$^{-1}$. The
radial velocities of the lines of the primary and the secondary
components amount to  \mbox{$-30$} and $+110$~km\,s$^{-1}$
respectively. The lower limit of the accuracy of finding $v\sin i$
of the cold component amounts to $15$~km\,s$^{-1}$. We estimate
the  accuracy of determining the radial velocities of the
components as   \mbox{$3$--$8$~km\,s$^{-1}$}.  However, if we
assume a more complex configuration of   stars in this system
(three stars or a star with a shell/magnetosphere),  all the
estimates of the parameters of the secondary component should be
considered as their upper value.

Quantitative estimates of the abundance of chemical elements in the atmospheres of the
  binary system components can be made only after the
accumulation of additional observational material and its
analysis. In a qualitative manner we can state the following about
the binary system: the main component is a chemically peculiar
star, the secondary is apparently the spectral type B9 star with a
normal chemical composition of the atmosphere. The abundance of
strengthened  Fe, Cr, Ti lines in the observed spectrum supports
this conclusion.  We have identified several Nd\,III lines.
Modeling of the silicon   lines even in the first approximation
detects significant discrepancies between the depth of the Si\,II
4621\,\AA\ and Si\,III 4552, 4567, 4574\,\AA{} lines. The
calculation of  theoretical spectra was based on the use of data
on the spectral lines from the VALD
database~\cite{vald:1,vald:2} and was carried
out using the {\tt Synthmag} code~\cite{kochukhov:2007}.
The magnitude of the microturbulence velocity was assumed to be
typical of CP stars within this temperature range:
\mbox{$\xi_\mathrm{micro}=1.5$~km\,s$^{-1}$}. It should be noted
that the rapid rotation of the secondary component greatly
simplifies the analysis of the magnetic field of the main star,
since the strongly broadened lines are practically invisible in
the
 combined spectrum.  A much greater contribution to the
approximation accuracy  of individual line profiles is made by the
magnetic field configuration. In our calculations the magnetic
field at the surface of the first star was supposed to amount to
$12$~kG, but we cannot exclude a more complex geometry which is
common in CP stars of the  B7--B9 spectral classes. In the latter
case the surface field of the star may prove to be larger. A
distortion in the shapes of profiles of most lines is noteworthy.
These distortions can be explained as an overlap of another
spectrum, close to the main component of the system in its
composition and effective temperature but with no radial velocity
shift.  We are confident that the discovered effect is not due to
the extraction errors but reflects the physical effect. A possible
explanation for all the listed facts is given in the next section.

\section{DISCUSSION}
The new magnetic star we have discovered,  HD\,34736, is
a unique object. Three measurements of polarized spectra show a
variable magnetic field, the magnitude of the longitudinal
component of which varies from 0 to almost \mbox{$-4500$}~G.
  Along with the magnetic field   we also
observe a spectral variability. The character of the latter
indicates the presence in the spectrum of the lines belonging to
the second star, somewhat cooler
\mbox{($T_\mathrm{eff}=11500$~K)}. The study of the HIPPARCOS
photometry data allows to select several possible periods of
brightness variability, where the most likely of them is
$0\fd3603$.  Apparently, this value is the orbital period of the
binary system. Radial velocities of the components,  obtained by
decomposing the spectrum into components, support this assumption.
All our results concerning the modeling of the spectrum of the
second star are preliminary. For a more detailed analysis new
observations are required in order to measure the radial velocity
of the system and the longitudinal magnetic field of the star. The
rotation period of the magnetic star in the HD\,34736 system can
only be determined after the analysis of a sufficient number of
measurements of its magnetic field. To make the final conclusion
on the presence of a magnetic field in the second star, circularly
polarized spectra with a very high signal-to-noise ratio are
required.

We  have tried to divide the spectrum of the star into the
components which allowed us to estimate the effective
temperatures,  surface gravity, and rotation velocities of both
components. We could obtain the preliminary value of \mbox{$v\sin
i=130$ km\,s$^{-1}$} for the secondary component only from one
line of Mg\,II~4481\,\AA, which manifests itself well in the right
wing of the line of the composite spectrum from \mbox{October 23,
2013.}  If we assume that the main contribution to the observed
spectrum is made by two stars with effective temperatures of
$T_\mathrm{eff}$  equal to $13\,700$ and $11\,500$~K, a larger
part of the observed hydrogen H$_\beta$ line can be described
well, except for its short-wavelength wing. At the same time, the
shapes of many lines reveal features that can be interpreted as
traces of the third spectrum with the chemical composition and
temperature of the main star.  The nature of this spectrum remains
to be clarified, but we can now confidently say that its
manifestation is not a result of processing errors: the short-wave
wing of the composite line H$_\beta$ can be described with high
accuracy only in the presence of the third spectrum. In our
opinion the physical explanation of such a complex spectrum may be
as follows. The star HD\,34736 represents a short-period binary
system, the main component of which, with a higher temperature, is
a magnetic star. The stars make one revolution around the common
center of gravity  in about $0\fd3603$. This results in
substantially different radial velocities of individual
components. The brightness curve of HD\,34736
(Fig.~\ref{figure:2}) has  a second minimum at phase
\mbox{$\varphi\approx0.73$}
 which can be caused by the spatial orientation of the system relative to the observer
and the dimension ratio of stars. It is very likely that
interchange of matter takes place in this system, forming a shell
or a tail. We detect the traces of this matter in form of the
third component of the spectrum. The presence in it of the same
lines as in the spectrum of the main star can tell of the origin
of   matter of this shell. The spectral range we observe bears no
prominent signs of emission. However,
Leone~\cite{leone:1994} states that HD\,34736 is a
confirmed  X-ray source. The X-ray activity in the case of
main-sequence B-type stars  is often a sign of either the presence
of a powerful magnetosphere or close interacting components. It is
possible that   HD\,34736 combines both cases. Therefore, the
further study of this star can be extremely important to address
the origin and evolution of close binary systems possessing a
strong magnetic field. But at the same time, we cannot exclude
other possible explanations of observations, since the total  data
set is still little.

{\bf Acknowledgments}\\
This work was partially supported by the Russian Foundation for
Basic Research (grant no.~\mbox{12-02-00009-a}). The
observations on the \mbox{6-meter} BTA telescope were conducted
with the financial support of the Ministry of Education and
Science of the Russian Federation (state
contracts no.~14.518.11.7070, 16.518.11.7073).

\begin{flushright}
{\it Translated by A.~Zyazeva}
\end{flushright}

\end{document}